# Making FPGAs accessible to Scientists and Engineers as Domain Expert Software Programmers with LabVIEW

Hugo A. Andrade
National Instruments
Berkeley, CA, USA
hugo.andrade@ni.com

Simon Hogg
National Instruments
Austin, TX, USA
simon.hogg@ni.com

Stephan Ahrends
National Instruments
Munich, Germany
stephan.ahrends@ni.com

*Abstract*— **In this paper we present a graphical programming framework, LabVIEW, and associated language and libraries, as well as programming techniques and patterns that we have found useful in making FPGAs accessible to scientists and engineers as domain expert software programmers.**

## I. INTRODUCTION

Many scientists and engineers do software programming as a (significant) part of their daily job. Either setting up experiments, taking measurements, developing prototypes, testing their designs, embedding software as part of larger systems, or modeling and analyzing the world around them. The type of software that they develop varies significantly with their domain or task that they are performing, and many times is categorized as scientific programming, embedded programming, test software development, and more recently cyber-physical systems programming, etc. Many of these scientist and engineers are domain experts in other fields like medicine, biology, chemistry, mechanical engineering, etc. and are not formally trained as computer scientists or electrical engineers, but have learned software programming well enough to be proficient in their own domain, and use software programming to their advantage. So having to go even further to understand how to program an FPGA is a foreign concept.

For such scientists and engineers having a programming framework that makes their programming job easier is a welcome help. In this paper we present one such framework, LabVIEW (Laboratory Virtual Instrument Engineering Workbench), specifically targeted to scientist and engineers, who need to program as part of their jobs. In addition to highlighting the merits of a visual framework and graphical programming language, we will also show the importance of intuitive and precise access to I/O, and a rich set of library components so that reuse can be exploited. We will also show how such framework has evolved from a programming environment to a system-level development framework for heterogeneous targets. Software programmers will find familiar elements in this flow, and will be eased into more hardware centric elements as needed/beneficial for specific requirements, in particular interfacing to real-world I/O at several levels.

The framework presents a platform-based design methodology, which supports multiple models of computation to capture the application at the right domain level, and then helps map that application into several pre-qualified platform targets. These heterogenous targets range from personal computers running desktop operating systems, to embedded real-time processors, to micro-controllers and FPGAs.

This paper is subdivided into the following sections: in Section II we describe the general LabVIEW framework in technical detail, including key language and environment features and the set of libraries supported. In section III we show how this environment can make development for software programmers trying to use FPGAs much easier. Section IV shows a supporting use case. Section V summarizes our conclusions and future direction.

## II. THE LABVIEW SOFTWARE DEVELOPMENT FRAMEWORK

LabVIEW is a graphical application programming environment originally developed in the mid 1980's by National Instruments Corporation for the Data Acquisition (DAQ), Test and Measurement (T&M) and the Industrial Automation (IA) markets. It is composed of several well integrated sub-tools targeted at making the development and prototyping of science and engineering applications that require interaction with real-world data and signals very simple and efficient. These sub-tools can be categorized as shown in Figure 1 [4], and are described in the following sub-sections.

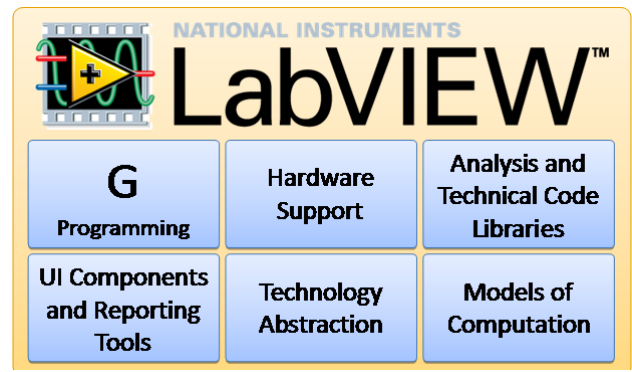

Fig. 1. LabVIEW Subtools by Category

*A. G Programming Language*

One of the key subtools is a compiler for the graphical programming language called *G*. G is a dataflow language that due to its intuitive graphical representation and programmatic syntax has been well accepted in the instrumentation industry, especially by scientists and engineers that are familiar with programming concepts but are not professional software developers and rather domain experts. Using it, they can quickly tie together data acquisition, analysis, and logical operations and understand how data is being modified. Though it is easy to use and flexible, it is built on an elegant and practical model of computation.

The idea behind G was to provide an intuitive flowchart-like block diagram view to the domain expert software programmer, since most scientists and engineers understood that concept; and it became the primary syntactical element in LabVIEW. The semantics follow homogeneous structured dynamic dataflow, which combines constructs from imperative and functional languages [1], where actor nodes (operations or functions called virtual instruments or "VI's") operate on data as soon as it becomes available, rather than in the sequential line-by-line manner that most programming languages employ. The VI's are connected via unidirectional edges (called "wires") as shown on Figure 2. VI's are either primitives built into G or sub-VI's written in the G language. Dataflow allows the user to not worry about memory allocation, garbage collection, and concurrency. The focus is on data, who produces and who consumes it, and what operation is performed on the data, rather than on how to pass data from one place to the other, or when the data arrive.

The user interface is presented through a "front panel" that provides "controls" and "indicators" through which the user sends and receives information. The programmatic hierarchical interface is specified through the connector pane.

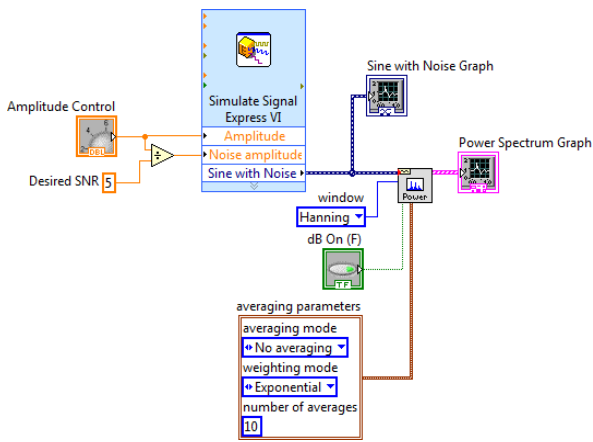

Fig. 2. LabVIEW Block Diagram

*B. Hardware Support*

Typically, integrating different hardware devices can be a major pain point when automating any test, measurement, or control system. Worse yet, not integrating the different hardware pieces leads to the hugely inefficient and error-prone process of manually taking individual measurements and then trying to correlate, process, and tabulate data by hand.

LabVIEW makes the process of integrating hardware much easier by using a consistent programming approach no matter what hardware is being used. The same initialize-configure-read/write-close pattern is repeated for a wide variety of hardware devices, data is always returned in a format compatible with the analysis and reporting functions, and the user is not forced to search through instrument programming manuals to find low-level message and register-based communication protocols unless specifically needed.

LabVIEW has freely available drivers for thousands of NI and third-party hardware devices (e.g. scientific instruments, data acquisition devices, sensors, cameras, motors and actuators, etc.) In the rare case that a LabVIEW driver does not already exist, the user has tools to create his own, reuse a DLL or other driver not related to LabVIEW, or use low-level communication mechanisms to operate hardware without a driver.

The cross-platform nature of LabVIEW also allows the user to deploy his code to many different computing platforms. In addition to the desktop OSs (Windows, OS X, and Linux), LabVIEW can target embedded real-time controllers, ARM microprocessors, and field-programmable gate arrays (FPGAs), so he can quickly prototype and deploy to the most appropriate hardware platform without having to learn new tool chains.

*C. Analysis and Technical Code Libraries*

LabVIEW tailors the G programming language to engineering and scientific use by incorporating hundreds of specialized functions and algorithms that are not typically included with general-purpose programming languages.

In addition to the standard programming language constructs, LabVIEW contains functions for:

- String, array, and waveform manipulation
- Signal processing, including filters, windowing, spectral analysis, and transforms
- Mathematical analysis, including curve fitting, statistics, differential equations, linear algebra, and interpolation
- Communication, including high-level protocols, HTTP, SMTP, FTP, TCP, UDP, Serial, and Bluetooth
- Report generation, file I/O, and database connectivity
- Add-on packages for specialized disciplines, such as:
    - Control design and simulation
    - Sound and vibration analysis
    - Machine vision and image processing
    - RF and communication

All of the included functions in LabVIEW work seamlessly with the data acquired from the supported hardware, and no special conversion or data movement setup is needed.

*D. UI Components and Reporting Tools*

Every LabVIEW block diagram also has an associated front panel, which is the user interface of the application. On the front panel generic controls and indicators such as strings, numbers, and buttons or technical controls and indicators such as graphs, charts, tables, thermometers, dials, and scales can be placed. These are designed for engineering use, meaning the user can enter SI units such as 4M instead of 4,000,000, change the scale, export data to tools such as NI DIAdem and Microsoft Excel, and they can be completely customized.

In addition to displaying data as an application is running, LabVIEW also contains several options for generating reports from test or acquired data. The user can send simple reports directly to a printer or HTML file, programmatically generate Microsoft Office documents, or integrate with NI DIAdem for more advanced reporting. Remote front panels and Web service support cab be used to publish data over the Internet with the built-in Web server.

*E. Technology Abstraction*

LabVIEW quickly adopts technology advances in personal and embedded computing in such a way that the domain expert user gets the new capabilities without having to learn significant new paradigms. Examples of this approach include how LabVIEW is able to automatically generate multithreaded code for execution on multicore processors or program FPGAs to gain the speed and reliability of custom hardware chips without the user needing to learn the underlying details of multithreading, or hardware description languages for FPGAs. The same applies to new OSs, networking protocols, etc.

*F. Models of Computation*

When LabVIEW was first released, G was the only way to define the user functionality. Much has changed since then. The user can now pick the most efficient approach to solve the problem at hand. Examine the following considerations:

- Graphical data flow is the default model of computation for LabVIEW

- Statecharts provide a higher level of abstraction for state-based (control) applications.

- Simulation diagrams are a familiar way of modeling and analyzing dynamic systems.

- Formula Nodes put simple mathematical formulas in line with the G code. MathScript is math-oriented, textual programming for LabVIEW that can be used to call .m files without the need for extra software.

- DLL calls, ActiveX/.NET communication, and the inline C node let the user reuse existing ANSI C/C++ code and code from other programming languages.

- CLIP and IP integration nodes import FPGA intellectual property based on VHDL or other HDLs.

- Single-cycle Timed Loops (SCTL) provide a synchronous-reactive model of computation for register-transfer level (RTL) hardware design.

These flexible models of computation allow the domain expert to pick the right tool for the particular problem being solved. In any given application the developer is likely to use more than one approach, and LabVIEW is a good tool to quickly tie everything together.

*G. Levels of Abstraction*

LabVIEW can express functionality at several levels that enable the user to trade off productivity or ease of use with performance of the resulting systems. It provides several ways to accomplish similar tasks, so the user can make balance simplicity and customization on a task-by-task basis.

Express VIs offer quick and easy configuration of VIs, but are somewhat limited in flexibility. Many of these types of VIs generate lower level G code that can be made accessible to the user as a starting point to develop more flexible code. Third party developers can themselves offer Express VIs to end-users.

Non-express VIs, where the Application Programming Interface (API) is exposed directly, are categorized as productive (high-level) or low-level. The most common way to program in LabVIEW is using high-level functions that strike a balance between abstracting the unnecessary administrative tasks such as memory management and format conversion, but keep the flexibility of being able to customize almost every aspect of whatever task you need to accomplish. In contrast, when one needs to be able to completely define every detail of a task, LabVIEW offers the same low-level access as available in traditional programming languages.

VIs at all levels of abstraction are provided as part of the main libraries, examples, additional libraries, third party libraries, and, in a growing number, by an open community. To deal with the large number of options a comprehensive VI search facility and hyper-linked documentation are provided in the framework.

III. ENABLING FPGAS FOR DOMAIN EXPERT SOFTWARE PROGRAMMERS

In this section we present some framework extensions (LabVIEW FPGA [6]) as well as techniques and patterns that focus on making FPGAs more accessible to LabVIEW software programmers, in particular those that are domain experts, focusing on protocol-aware testing, software-defined instruments, and embedded and cyber-physical systems.

*A. Integrated Synthesis Tools*

The Xilinx synthesis tools are integrated with LabVIEW FPGA. They are invoked automatically on hidden generated code to create a bitfile to download and run on the FPGA.

*B. Reference Target Platform*

In order to make targeting FPGAs more consistent, we defined a canonical FPGA platform that is not limited to just a fabric, but included an instruction processor and attached I/O. It centered on the common use case of real-world I/O access, and was specialized for an FPGA-based computing platform. We call it *Reconfigurable I/O (RIO)* [5], as shown in Figure 3.

There are three main computing engines: an externally attached or internal instruction processor, typically running a real-time operating system, an (optional) remote host computer, typically running a desktop operating system, and the FPGA fabric. All of these computing engines are programmable by LabVIEW. Depending on the application and the knowledge or experience of the programmer, parts of the code may run on the host computer, the real-time computer or the FPGA. The framework attempts to show a consistent programming environment, yet has not hidden the boundaries between these components.

I/O attached to the FPGA is viewed as part of a RIO target, and has first class representation within the LabVIEW FPGA environment. The nodes present an interface to the program in terms closest to the domain expert, e.g. analog input in engineering units, as opposed to digital protocols to specific analog to digital converters. Similarly control of the timing of the critical I/O operations is done in high-level terms.

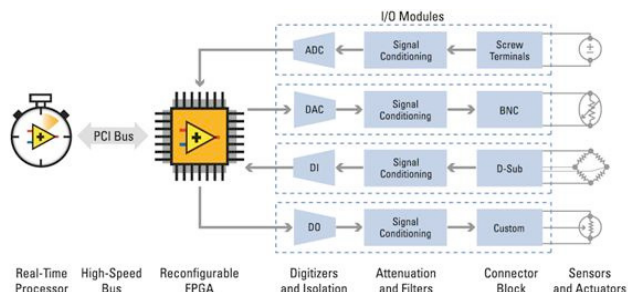

Fig. 3. LabVIEW RIO Architecture

### C. Pre-configured FPGA bit personalities with I/O

As we have mentioned, access to real-world I/O is one of most important requirements for the type of domain experts that we are targeting. Yet, many of them do not want to program an FPGA just to have synchronous deterministic access to I/O. The *NI Scan Engine* [3] enables efficient access to coherent sets of I/O channels attached to the FPGA, using a scan that stores data in a global memory map and updates all values at a single rate, known as the scan period. To do this a specialized IP is placed on the FPGA that interfaces to I/O modules connected to the FPGA. Yet, at the same time, it allows the rest of the fabric to be available for further FPGA computing or connecting to other types of specialized I/O that may not only interface to the Scan Engine.

### D. FPGA Extensions

Whenever possible we try to make I/O functionality directly available to the instruction processor, either by using the Scan Engine or other "relaying" techniques. This gives the programmer the most flexibility. In doing so, we make choices on levels of API supported. For example, in the case of RIO-enabled RF instruments, we have pre-built bit personalities onto the FPGA that allow the board to appear as a more traditional instrument, by supporting standard APIs for that class of instrument. Yet, domain experts want to make changes and specialize such a software-defined instrument. Such a change in the FPGAs makes it difficult to preserve the investment of code that has been made on the host already, which uses the standard API; matching the original validated FPGA design as well. So a set of side-band interfaces were defined that allow most of the matched FPGA-Host code to continue to operate as originally intended, but allowing for extensibility. The side-band interface includes feature or signature export from the FPGA and import on the host, as well as communication through host-FPGA boundary.

### E. Effective Use of Hardware-Software Communication

In order to be able to start with a design that has most of the code on the instruction processor, and migrates only those part needed to the fabric, we need to minimize the development and run-time cost of doing so. This means that the inter-process communication has to be independent of target and has to be efficient. The LabVIEW RIO framework supports both queue and memory register abstractions to communicate between tasks on the same processor or FPGA, as well as across these entities. If the entities are across a bus, it automatically instantiates DMA channels and engines to support the operation.

### F. Models of Computation

Different jobs require different tools. Similarly, depending on the tasks, the user may want to code application components using different models of computation. Just like in the general framework, multiple models of computation are available for the FPGA. In particular, the general G structured homogeneous dataflow is the default model of computation. The Single Cycle Timed Loop sub-framework discussed below follows a synchronous reactive model of computation consistent with execution on clock ticks. For control applications we also offer a modal model that follows StateCharts syntax and semantics, including hierarchical and concurrent states. Some models of computation like continuous time (mainly for physical subsystem modeling) are not available for implementation on FPGA yet, but are available for co-simulation.

### G. Consistent IP and Programming Model

As part of making software components portable between targets it is important to not only have a consistent programming language, but also consistent set of libraries that port across most targets. Each of these libraries is optimized and the resource utilization is made clear, either in documentation or as part of the framework information system.

### H. Optimization

#### 1) Algorithmic IP and High-level Synthesis

Optimizing an algorithm for an FPGA can be a complex task since the concurrency and connectivity options are fairly large. As an add-on to the LabVIEW FPGA Module, LabVIEW FPGA IP Builder generates high-performance FPGA IP by leveraging high-level synthesis (HLS) technology. HLS generates efficient hardware designs from three components: 1. an algorithm implemented in a high-level language such as LabVIEW, 2. visual synthesis directives specifying the performance and resource objectives, and 3. an FPGA target with known characteristics. It can be used to automatically optimize FPGA VIs and easily port desktop code to the FPGA. The user is able to quickly explore designs with rapid performance and resource estimates. He can reuse IP, unmodified, to adapt to different application requirements.

*2) Low-level IP and Graphical RTL Abstraction*

Sometimes in order to optimize an algorithm or to connect to detailed I/O protocols it is necessary to describe the operations at the cycle or register transfer level (RTL). LabVIEW provides the Single Cycle Timed Loop model of computation in which the user, when needed, can describe the operations a cycle at a time. Within this environment the compiler takes advantage of the fact that data needs to only propagate data between adjacent flip-flops and removes part of the logic that would normally ensure that the regular dataflow semantics are followed between VIs.

*I. External IP*

Even if we provide much IP at the high-level, sometimes the end-user either has legacy IP or has available to them IP written in a different programming or hardware description languages. In such a case, it is important to be able to import that IP, and make it appear in a consistent and native form to the LabVIEW framework. CLIP (Component Level IP) and IPIN (IP Integration Nodes) are two ways to import external IP into LabVIEW FPGA. A CLIP Node executes independently and in parallel to the IP developed in LabVIEW FPGA. In addition, CLIP can interface directly with the FPGA clocks and I/O pins. In contrast, the IP Integration Node is inserted into the LabVIEW FPGA block diagram and executes as defined by the dataflow of the LabVIEW VI. As part of the LabVIEW dataflow execution, the IP integration Node provides the ability to verify the overall application behavior and timing using the cycle-accurate simulation tools.

*J. Soft Computing Targets*

FPGAs are very flexible, including the development of instruction processors on the fabric. These processors are complete but are not the most powerful, yet the fact that the user can dedicate a processor to a specific I/O component is very useful. Using some of the IP integration mechanisms described above, we have been able to integrate popular soft processors such as MicroBlaze, by using the Xilinx tools to define the processor and its I/O. The I/O then appears as connection to the rest of the LabVIEW FPGA code for setup in accelerated computing or directly attached to I/O. The processor can be programmed today in C, and resulting compiled code can be overlaid onto an existing bit file without recompiling the rest of the fabric portion.

*K. Consistent Debugging Techniques and Simulation*

In addition to consistent models of computation, libraries and patterns, it is important to have a consistent debugging and simulation flow in between targets as well. In the LabVIEW framework a developer can start development on the host, and can complete all the logic or algorithm development there, then they can move all to the real-time processor in case they need to directly access I/O, and eventually to the LabVIEW FPGA. Once on the FPGA they can see a cycle accurate simulation of the application, and only then would they need to target the actual FPGA, so that the compilation time is minimized.

*L. High-level Language Features*

Software engineering principles like object-oriented design are very useful to domain experts as well, especially as their code base grows incrementally or when they are starting on large projects to begin with. These techniques are also available in graphical form with LabVIEW and they extend to LabVIEW FPGA as well.

IV. USE CASE

To demonstrate some of the framework features and recommended techniques listed in sections II and III, we describe in this section how we developed a device that connects to a regular pair of analog speakers and enables them to play music from iOS devices over the air. To this effect we used the LabVIEW framework and a RIO board to develop an AirPlay-based [7] Wireless Music System (WMS).

Since we wanted for this design to be available as reference for students, we selected as a base interface board the NI myRIO [2], which is an embedded hardware device that introduces students to industry proven RIO technology and allows them to quickly design real, complex engineering systems. In our case, it provides the following platform elements needed for our implementation: 1. Wi-Fi access for over the air music delivery; 2. a Digital to Analog Converter (DAC), connected to audio quality analog output (for analog speakers); and 3. a Xilinx Zynq 7010 all-programmable SoC, implementing the basic RIO architecture elements: a dual-core ARM processor to run the basic protocols, and high-level soft real-time processing; a reconfigurable fabric to interface to the DAC, provide low-level timing, and hard real-time processing.

There are a couple of free implementations of the AirPlay protocol [7] for ARM processors running Linux, which we were able to leverage. The main application is developed using LabVIEW RT (Figure 4). It spawns a process running ShairPort, which implements basic protocols and decoding for the music transmitted from iOS devices. ShairPort was compiled from C code available online [7]. To advertise AirPlay service ShairPort searches for an installed Avahi mDNS service (binary opkg) and connects to it directly.

The ShairPort process provides the data in lossless format to the main LabVIEW RT processing loop via UNIX pipes. The main loop implements a simple equalizer using a 3-channel filter bank (Figure 5), using the built-in libraries. It then passes the data to a Host-FPGA DMA FIFO queue, so that we can implement on the FPGA the only part that really requires very precise timing, namely the 44KHz interface to the analog out port (Figure 6). Notice that as domain experts we do not need to worry about the details of the interface to DAC, but just provide high-level data to the I/O interface. Similarly we described the timing as a simple loop timer that gets the specific timing requirements from the RT processor via a run-time parameter.

The prototype is shown in Figure 7, with the myRIO device attached to the speakers, streaming music from an iPhone.

## V. CONCLUSIONS AND FUTURE DIRECTIONS

In this paper we have presented a graphical programming framework, LabVIEW, and associated language and libraries, as well as programming techniques and patterns that we have found useful in making FPGAs accessible to scientists and engineers as domain expert software programmers.

We focused on a single RIO module as target, i.e. one instruction processor, attached fabric and I/O, and we only touched briefly on some of the challenges of networking RIO subsystems. In the future, we would like to focus on networks of RIO modules, and some of the challenges of programming such a system. In particular, we would like view the entire system of processors and FPGA as one large target, i.e. system-level synthesis.

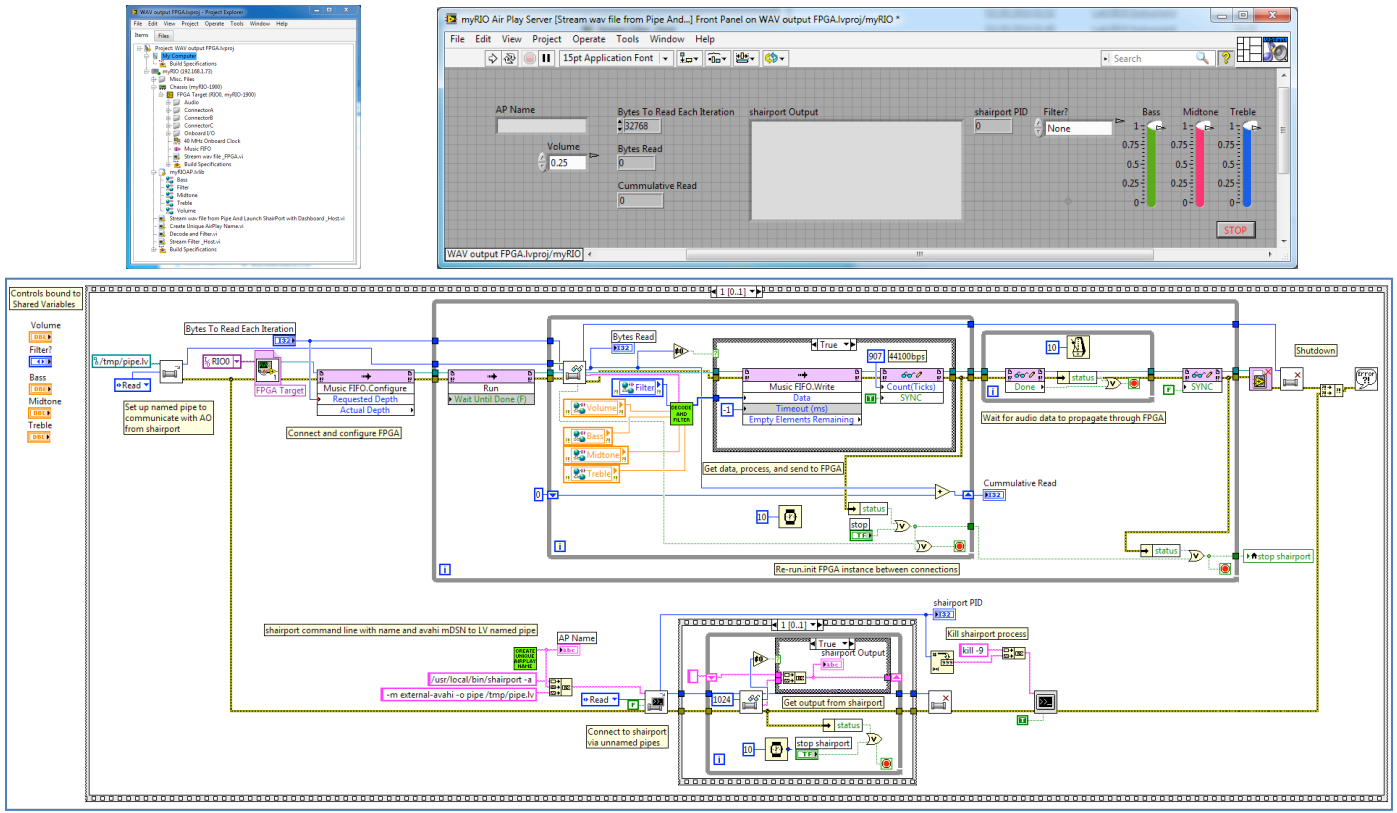

Fig. 4. LabVIEW Project Explorer View (top left) and LabVIEW Linux RT Top-level App Front Panel (top right) and Block Diagram (bottom)

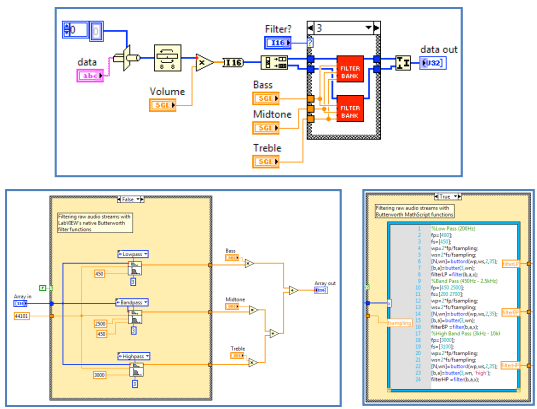

Fig. 5. LabVIEW Linux RT Decoding and Processing BDs

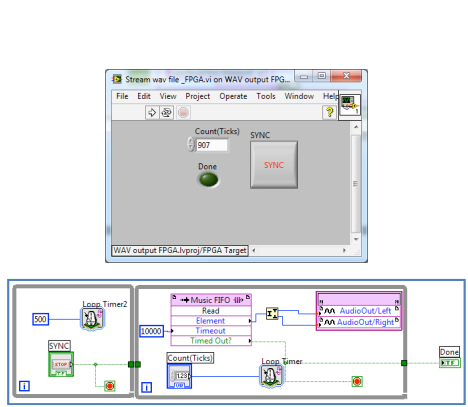

Fig. 6. LabVIEW FPGA DAC Timing FP and BD

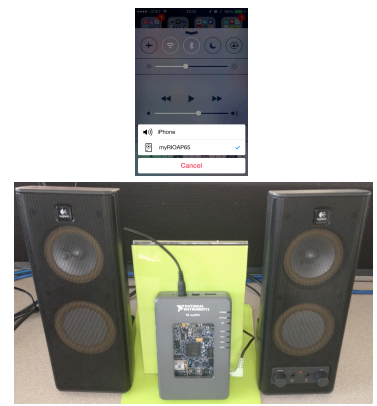

Fig. 7. myRIO WMS Protoytpe

14